\documentclass[preprint, superscriptaddress, longbibliography]{revtex4-1}

\newcommand{\bl}{\begin{flalign}}
\newcommand{\enl}{\end{flalign}}

\newcommand{\mc}[1]{\mathcal{#1}}

\newcommand{\tdse}{time dependent Schr\"{o}dinger equation}

\newcommand{\eq}[1]{Eq. \eqref{#1}}

\newcommand{\half}{\frac{1}{2}}
\newcommand{\intf}{\int_{-\infty}^{\infty}}

\newcommand{\tr}{\text{Tr}}

\newcommand{\proj}[1]{\ket{#1}\bra{#1}}
\newcommand{\fig}[1]{Fig. \ref{#1}}

\usepackage{graphicx}
\usepackage{amssymb}
\usepackage{epstopdf}
\usepackage{amsmath,dsfont}
\usepackage{bm}
\usepackage{braket}
\usepackage{tikz}

\newcommand{\be}{\begin{equation}}
\newcommand{\ee}{\end{equation}} 
\newcommand{\bea}{\begin{eqnarray}}
\newcommand{\eea}{\end{eqnarray}}
 
\newcommand{\ba}{\begin{array}}
	\newcommand{\ea}{\end{array}}

\renewcommand{\Re}{\operatorname{Re}}
\renewcommand{\Im}{\operatorname{Im}}

\newcommand{\tord}{{\mathcal{T}}} 
\newcommand{\tordc}{{\mathcal{T}_\mathcal{C}}}  
\newcommand{\intc}{\int_\mc{C}}  
\newcommand{\dt}{\frac{d}{dt}}

\begin{document}
\title{When can quantum decoherence be mimicked by classical noise?}
	\author{Bing Gu} 

\altaffiliation[Present address:]{ Department of Chemistry, University of California, Irvine, CA 92697, USA}
\affiliation{Department of Chemistry, University of Rochester, Rochester NY 14627, USA}
\author{Ignacio Franco}
\affiliation{Department of Chemistry, University of Rochester, Rochester NY 14627, USA}
\affiliation{Department of Physics, University of Rochester, Rochester NY 14627, USA}
\date{\today} 

\begin{abstract}
Quantum decoherence arises due to uncontrollable entanglement between a system with its environment. However the effects of decoherence are often thought of and modeled through a simpler picture in which the role of the environment is to introduce classical noise in the system's degrees of freedom. Here we establish necessary conditions that the classical noise models need to satisfy to quantitatively model the decoherence. Specifically, for pure-dephasing processes we identify well-defined statistical properties for the noise that are determined by the quantum  many-point time correlation function of the environmental operators that enter into the system-bath interaction. In particular, for the exemplifying spin-boson problem with a Lorentz-Drude spectral density we show that the high-temperature quantum decoherence is quantitatively mimicked by colored Gaussian noise. In turn, for dissipative environments we show that classical noise models cannot describe decoherence effects due to spontaneous emission induced by a dissipative environment. These developments provide a rigorous platform to assess the validity of classical noise models of decoherence. 
\end{abstract}

\maketitle

\section{Introduction}
The inevitable interaction between a quantum system with its surrounding environment leads to decoherence \cite{Breuer2002, Schlosshauer2007, Gu2017, Gu2018, Gu2018c, Hu2018}.  The decoherence  occurs because such interaction leads to system-bath entanglement that turns a pure system state to a statistical mixture
of states. Understanding quantum decoherence  is important for a wide range of fields such as quantum computation and quantum information processing \cite{Nielsen2011}, quantum control\cite{Shapiro2003}, measurement theory, spectroscopy, molecular structure and dynamics \cite{Valkunas2013}.

There are several theoretical frameworks to understand quantum decoherence and the effective dynamics of open quantum systems \cite{Breuer2002}. The most rigorous one 
of them consists of \emph{explicitly} solving the time-dependent Schr\"{o}dinger equation for the system and its environment and then tracing out the environmental degrees of freedom to obtain the system's reduced density matrix.  However, this approach, while desirable \cite{Hu2018, Franco2008_charge, Franco2008}, is often intractable due to the exponentially increasing computational cost of solving the time-dependent Schr\"{o}dinger equation with system/environment size. This limitation has lead to significant advances developing methods in which the effect of the bath is considered \emph{implicitly}  \cite{Breuer2002, Gu2017a} 
such as perturbative quantum master equations \cite{Tanimura2006}, path integral
techniques \cite{Walters2015} and hierarchical equations of motion
\cite{Tanimura2012, Tanimura1989}. Despite this important progress,  following the reduced
dynamics of a primary system of interest interacting
with a general quantum environment remains an outstanding challenge.

Due to the conceptual and technical complexities in dealing with the  system plus environment fully quantum mechanically, an alternative approach  is to simply consider that the effect of the environment is to introduce classical noise in the system's degrees of freedom \cite{Budini2001, Yang2017, Stern1990, Kubo1969, Gelzinis2015, Leon-Montiel2013, Costa-Filho2017, Chenu2017, Spanner2009}. In this picture, quantum dissipation is  mimicked by stochastic terms in the equation of motion that introduce random transitions between system energy eigenstates. In turn, pure-dephasing processes are modeled by introducing dynamic disorder (or, equivalently, spectral diffusion) in which classical noise perturbs the energy of the system eigenstates leading to an accumulated random phase. Decoherence arises by averaging over an ensemble of these stochastic but unitary quantum dynamics.  

Note that this implementation of decoherence through noise requires averaging over an ensemble of realizations each one evolving unitarily. The corresponding ensemble of unitary evolutions represents a nonunitary evolution of the density matrix of the system. By contrast, ``true" decoherence occurs for a single-quantum system that becomes entangled with environmental degrees of freedom. The unitary deterministic evolution of the system plus environment leads to a nonunitary evolution of the reduced density matrix of the system. This conceptual difference between noise and true decoherence is known~\cite{Schlosshauer2007, Joos2013}. However, unless this difference is probed explicitly, the noise model can mimic well the effects of decoherence since they both effectively lead to a damping of  coherences.  In fact, this stochastic  picture with classical noise has been widely used in chemistry and physics to capture the 
 loss of interference \cite{Stern1990, Spanner2009}, optical line shapes \cite{Kubo1969, Gelzinis2015}, noise-assisted energy transport \cite{Leon-Montiel2013}, non-Markovian dynamics \cite{Costa-Filho2017}, Landau-Zener \cite{Wubs2006, Kayanuma1985} and central-spin problems \cite{Yang2017} and in  the quantum simulation of open many-body systems \cite{Chenu2017}.

The fundamental question that  arises in this context is what is the regime   of validity and the limitations of the classical noise picture. 
An initial discussion of this problem was provided by Stern \emph{et al.} \cite{Stern1990} where it is argued that the loss of quantum interference can be mimicked by the phase uncertainty introduced by the classical noise. However, no formal criteria for the validity of classical noise was provided. Here we  identify  necessary conditions under which the decoherence effects induced by a quantum environment in a quantum system can be  understood and modeled through classical noise. 
Such conditions are obtained by comparing the  reduced dynamics of an open quantum system to the ensemble average of a series of unitary quantum trajectories generated by a stochastic Hamiltonian.  
We consider the effects of dissipation and pure dephasing independently and do not take into account their possible interference which was recently demonstrated in Ref. \cite{Gu2018}.    

This paper is organized as follows. 
Section \ref{sec:pure} introduces decoherence functions that arise due to system-bath entanglement and due to classical noise in the pure dephasing limit. 
Through a term-by-term comparison of their cumulant expansion, we isolate conditions on the classical noise that need to be satisfied to mimic the quantum dynamics. 
 These conditions are determined by the many-point time correlation functions of the
environment operators that enter into the system-bath interaction. 
The application of these conditions to the spin-boson model show that the decoherence effects  can be captured through colored Gaussian
noise provided that the environment time-correlation function  can
be described by a set of exponentially decaying functions. In turn, Sec. \ref{sec:diss} focuses on decoherence through quantum relaxation. We show that classical noise cannot describe decoherence induced by spontaneous emission and thus that these models are of limited applicability when spontaneous fluctuations play a critical role.

\section{Pure dephasing dynamics} \label{sec:pure}

 We  first focus on pure dephasing dynamics and establish  general criteria that needs to be satisfied to employ classical noise to mimic quantum decoherence. 
Pure dephasing  refers to a process in which the decoherence arises without energy transfer between system and environment. 
For a general composite system with Hamiltonian,
\be
H = H_\text{S} + H_\text{B} + H_\text{SB}
\ee 
where $H_\text{S}$  is the Hamiltonian of the quantum system, $H_\text{B}$  of
the environment and $H_\text{SB}$  the interaction between system and
bath, the pure-dephasing condition arises when $[H_\text{S}, H_\text{SB}]=0$. Even
 when this condition is not strictly satisfied, the pure-dephasing effects may still be the dominant effect when the environment dynamics is non-resonant with the transition frequencies of the system such that the dissipation is much slower compared to pure-dephasing effects. 
For this reason, the pure-dephasing limit has been useful in describing electronic decoherence in molecules \cite{Gu2018, Hu2018}, elastic electron-phonon interaction in solid state systems, loss of quantum interference \cite{Stern1990}, line shape in spectroscopic measurements \cite{Kubo1969}, 
vibrational dephasing in solvents \cite{Joutsuka2016} and the central spin problem \cite{Yang2017}.

Below we define decoherence functions that arise from system-bath entanglement and from noise-induced pure dephasing. By contrasting them we isolate conditions that the classical noise needs to satisfy to mimic the quantum decoherence. 

\subsection{ Quantum decoherence function}

For pure-dephasing dynamics, the system-bath interaction can be written as 
\be H_\text{SB} = \sum_\alpha  \ket{\alpha}\bra{\alpha} \otimes B_\alpha
\label{eq:hsb}
 \ee    
where $\{ \ket{\alpha}\}$ are the eigenstates of $H_\text{S}$ and $B_\alpha$ is a bath operator. 
Here we assume that the system and bath are uncorrelated at initial time such that the  density matrix can be written as 
\be \rho(0) = \rho_\text{S}(0) \otimes \rho_\text{B}(0), 
\label{eq:init}
\ee
where $\rho_\text{S}$ is the reduced density matrix for the system and $\rho_\text{B}$ for the bath.   
The Liouville-von Neumann (LvN) equation  in the interaction picture of $H_0 = H_\text{S} + H_\text{B}$ reads 
 \be i \frac{d}{dt} {\tilde{\rho}}(t) = [\tilde{H}_\text{SB}(t), \tilde{\rho}(t)] , 
 \label{eq:113}
 \ee 
 where $\tilde{A}(t) = U_0^\dag(t) A U_0(t)$ is the operator $A$ in this interaction picture and $U_0(t) = e^{-i H_0 t}$.
 For notational convenience, for system operators  $\tilde{A}_\text{S}(t) \equiv U_\text{S}^\dagger A_\text{S}U_\text{S} $ where $U_\text{S}=e^{-iH_\text{S} t}$. Similarly, for bath operators $\tilde{A}_\text{B}(t) \equiv U_\text{B}^\dagger A_\text{B}U_\text{B} $ where $U_\text{B}=e^{-iH_\text{B} t}$.
  Here and throughout we employ atomic units where $\hbar=1$.   
The solution to the LvN equation can be written as 
\be 
\tilde{\rho}(t) = \tilde{U}(t) \rho(0) \tilde{U}^\dag(t) 
\label{eq:114}
\ee 
where 
$ \tilde{U}(t) = \tord e^{-i \int_0^t \tilde{H}_\text{SB}(t')\,dt'}$ is the propagator in the interaction picture and $\tord$ is the time-ordering  operator.   
Using \eq{eq:hsb}, it follows that 
$ \tilde{H}_\text{SB}(t) =  \sum_{\alpha} \proj{\alpha} \otimes \tilde{B}_\alpha(t) $
and 
\be 
\begin{split} 
\tilde{U}(t) &= \tord \sum_{n=0}^\infty \frac{(-i)^n}{n!} \left( \int_0^t dt' \sum_{\alpha} \proj{\alpha} \otimes \tilde{B}_\alpha(t')\right)^n \\ 
&= \sum_{\alpha}   \proj{\alpha} \otimes \tord \sum_{n=0}^\infty \frac{(-i)^n}{n!}  \left( \int_0^t dt' \tilde{B}_\alpha(t')\right)^n = \sum_{\alpha} \proj{\alpha}  \otimes V_\alpha(t)
\end{split}
\label{eq:118}
\ee 
where $V_\alpha(t) \equiv \tord \exp\left(-i \int_0^t \tilde{B}_\alpha(t')\,dt' \right)$.
Inserting \eq{eq:118} into \eq{eq:114}, taking into account the uncorrelated initial system-bath state in \eq{eq:init},  
and tracing out the bath degrees of freedom (which is denoted by $\tr_\text{B}[\cdots]$) yields  the reduced density matrix for the system 
\be \tilde{\rho}^{\text{S}}_{\alpha \beta}(t)  = \bra{\alpha} \textrm{Tr}_\text{B}[ \tilde{\rho}(t)]\ket{\beta} 
 = {\rho}^{\text{S}}_{\alpha \beta}(0) \Phi_{\alpha \beta}(t).
\label{eq:solq}
 \ee 
Here 
\be \Phi_{\alpha \beta}(t) \equiv \tr_\text{B}[\rho_\text{B}(0) V^\dag_\beta(t)V_\alpha(t)] = \Braket{V^\dag_\beta(t) V_\alpha(t) },
\label{eq:df_def}
\ee  
 is the quantum decoherence function (QDF), which characterizes the decoherence effects for pure-dephasing dynamics.  
In this pure-dephasing dynamics, the diagonal matrix elements of the reduced density matrix representing populations in the energy eigenstates are not influenced by the environment as $\braket{ V^\dag_\alpha(t)V_\alpha(t)} = 1$. However, the off-diagonal elements of the density matrix decay with a rate determined by $\Phi_{\alpha\beta}(t)$. 

 If the initial state of the environment is pure, i.e., $\rho_\text{B}(0) = \ket{\chi}\bra{\chi}$, the QDF becomes 
\be \Phi_{ \alpha \beta}(t) = \braket{\chi | V^\dag_\beta(t) V_\alpha(t)|\chi }. 
\label{eq:115}
\ee 
 In this case, the absolute square of  decoherence function $|\Phi_{ \alpha \beta}|^2$  is known as the Loschmidt echo  $L(t)$ \cite{Goussev2012}. The Loschmidt echo measures the stiffness of the environment to the perturbation by the system and is deeply connected to quantum decoherence  \cite{Cucchietti2003}. A particular interesting case  is that for a two-level system with a initial state $\ket{\psi_0} = c_0\ket{0} + c_1\ket{1}$ , the Loschmidt echo connects directly to the purity of the system, defined as $\mc{P}(t) = \tr_\text{S} [\rho_\text{S}^2(t)] $, with the following relationship 
 \be \mathcal{P}(t) = 1 + 2|c_0|^2|c_1|^2 (L_{01}(t)-1) .\ee

\subsection{Noise-induced decoherence function}
Consider now a quantum system that is subject to classical noise. The noise is supposed to cause spectral diffusion, i.e. to introduce stochastic dynamics to the energy eigenvalues of the system. The effective Hamiltonian of the system for a particular realization of the noise is
\be H(t) = H_\text{S}  + \sum_\alpha \eta_\alpha(t) \ket{\alpha}\bra{\alpha} \ee 
where 
$\{\eta_\alpha(t)\}$ are real stochastic processes. For the Hamiltonian to be Hermitian the $\eta_\alpha(t)$ must be real.   
The density matrix for a single realization of the noise can be obtained from the LvN equation in the interaction picture of $H_\text{S}$ 
to yield 
\be i \frac{d}{dt} \tilde{\rho}_{\alpha \beta}(t)= (\eta_\alpha(t) - \eta_\beta(t)) \tilde{\rho}_{\alpha \beta}(t).
\label{eq:116}
\ee 
 Taking a statistical average of the solution of \eq{eq:116}  yields 
\be \overline{\tilde{\rho}_{\alpha \beta}(t)} = \Phi^\text{noise}_{\alpha \beta}(t) \rho_{\alpha \beta}(0),
\label{eq:solc}
\ee 
where we have introduced the noise-induced decoherence function (NIDF)
\be \Phi^\text{noise}_{\alpha \beta}(t) = \overline{e^{-i
		\int_0^t \Delta_{\alpha \beta}(s) \,ds }}, 
\ee 
$\Delta_{\alpha \beta}(s) \equiv \eta_\alpha(s) - \eta_\beta(s)$ and the overline denotes statistical averaging.

\subsection{Contrasting quantum and noise-induced decoherence functions}

Comparing Eqs. \eqref{eq:solq} and \eqref{eq:solc}, it is clear that
if the classical decoherence function coincides with the  quantum decoherence function, i.e., 
\be \Phi_{\alpha \beta}(t) = \Phi^\text{noise}_{\alpha \beta}(t) \qquad \forall \, \alpha, \beta, 
\label{eq:cond}
\ee 
  the noise picture of decoherence accurately mimics the entanglement process that leads to the decoherence. This formal relation offers a general structure to understand how classical noise models can be related to physical pure dephasing processes.  However, it does not offer a practical prescription to relate the decoherence dynamics with  the statistical properties of the noise as the quantum decoherence function involves two time-ordered exponentials of the bath operators which are generally not available. 

To make further progress, below we introduce a useful  operatorial identity for products of time-ordered exponentials and use it to develop a cumulant expansion of the quantum decoherence function.

\subsubsection{A useful operatorial identity}

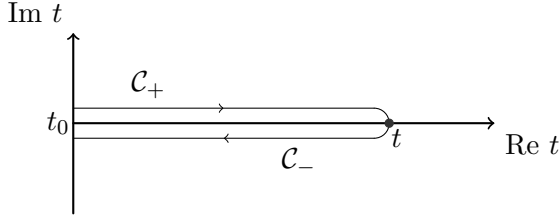
\begin{figure}[bt] 
	\begin{tikzpicture}[
	scale=2,
	line cap=round,
	dec/.style args={#1#2}{
		decoration={markings, mark=at position #1 with {#2}},
		postaction={decorate}
	}
	]
	\draw[thick,->] (0,0) -- (2.8,0) node[anchor=north west] {Re $t$};
	\draw[thick,->] (0,-0.6) -- (0,0.6) node[anchor=south east] {Im $t$};	
	\draw [->](0,0.1)--node[above,black]{$\mathcal{C}_+$} (1.0,0.1); \draw (1,0.1)-- (2,0.1);
	\draw [->](2.0,-0.1)--node[below,black]{$\mathcal{C}_-$} (1.0,-0.1) ; \draw (1,-0.1)-- (0,-0.1);
	\draw (2,0.1) arc (90:-90:1mm);
	
	\node at (2.1,0.0) [circle, scale=0.3, draw=black!80,fill=black!80] {};
	\node  at (2.15,-0.1) {$t$};
	\node at (-0.1,-0) {$t_0$};
	\end{tikzpicture}
	\caption{The complex time contour that is used in \eq{eq:identity}.}
	\label{fig:kc} 
\end{figure}

We now show that given two general Hermitian operators $A(t)$ and $B(t)$ 
	\be \bar{\tord} e^{i \int_0^t B(\tau)\,d\tau} \tord e^{-i \int_0^t A(\tau)\,d\tau} =  \tord_\mc{C} e^{-i \int_0^t (A(\tau_+) - B(\tau_-) )\,d\tau} 
	\label{eq:identity}
	\ee
	where $\bar{\tord}$ is the anti-chronological time-ordering operator,  and $\tord_\mc{C}$ is the contour-ordering operator defined in a complex time contour $\mc{C}$ as specified in Fig.~\ref{fig:kc}.
	The anti-chronological time ordering operator rearranges earlier-time terms to the left of the later-time ones, and the contour-ordering operator rearranges earlier-in-contour terms to the right of the later-in-contour ones. 
	  This contour consists of two time branches, the upper branch going forward in time from $t_0+i\epsilon  \rightarrow t+i\epsilon $ and the lower one going backward in time from $t-i\epsilon \rightarrow t_0-i\epsilon$ where $\epsilon = 0^+$ is an infinitesimal positive number. 

Equation \eqref{eq:identity} can be understood as a direction extension of the semigroup property of the evolution operator [$U(t,t') = U(t,t'')U(t'',t')$] from real time to a complex time contour.  
 A formal proof is provided as follows. 
We first note that 
  \be 
   \bar{\tord} e^{i \int_0^t B(\tau)\,d\tau} \tord e^{-i \int_0^t A(\tau)\,d\tau} =
  \tordc e^{i \int_0^t B(\tau_-)\,d\tau } e^{-i \int_0^t A(\tau_+)\,d\tau}    
  \label{eq:121}
  \ee
  due to the fact that the effects of the two time-ordering operators in the left-hand side are being taken care of by the contour-ordering operator. 
  Here the subindex $ \pm $ indicates the upper/lower time branch of the  contour. 
Using the Baker-Campbell-Hausdorff formula \cite{Rossmann2002} $e^X e^Y = e^{X+Y+\half [X,Y] + \cdots }$  yields 
\begin{widetext}
\be 
 \tordc e^{i \int_0^t B(\tau_-)\,d\tau } e^{-i \int_0^t A(\tau_+)\,d\tau}   
= \tordc \exp \left\{ i\int_0^t( B(\tau_-) - A(\tau_+))d\tau - \frac{i^2}{2} \iint_0^t  [ {B}(\tau_-),  A(\tau'_+)]\,d\tau d\tau' + \cdots \right\} 
\label{eq:tmp}
\ee
\end{widetext}
Now, commutators vanish under the contour-ordering operator 
\be \tordc \{[A(\tau),B(\tau')] \}= \tordc \{A(\tau)B(\tau') - B(\tau')A(\tau)\} = 0 
\ee 
as the two terms will be ordered in the same way by the contour-ordering operator.
Then all commutators and nested commutators in \eq{eq:tmp} vanish, yielding the identity in \eq{eq:identity}.  

The utility of \eq{eq:identity} is that it enables us to express the two time ordered exponentials in $\Phi_{\alpha\beta}(t)$ in terms of a single contour-ordered exponential. As shown below, such exponential admits a simple cumulant expansion that will enable us to connect the desirable statistical properties of the noise with quantum time-correlation functions.

\subsubsection{Decoherence function in the contour}

Using \eq{eq:identity} it follows that 
\be V_\beta^\dag(t)V_\alpha(t) = 
 \tordc \exp\left(i\int_0^t (\tilde{B}_\beta(\tau_-) - \tilde{B}_\alpha(\tau_+))\,d\tau 
\right). 
\label{eq:112}
 \ee 
This equation can be simplified further if we define a function in the contour as  
\be \mc{B}_{\alpha \beta}(\tau) = 	\theta_\mc{C}(t-\tau) \tilde{B}_\alpha(\tau)  + 	\theta_\mc{C}(\tau - t) \tilde{B}_\beta(\tau) 
\ee
where $\theta_\mc{C}(\tau-\tau')$ is the Heaviside step function defined in the contour,  $\theta_\mc{C}(\tau-\tau') = 1$ if $\tau$ is later than $\tau'$ in the contour and $\theta_\mc{C}(\tau-\tau') = 0$ otherwise. 
Using this definition, Eqs. \eqref{eq:112} and \eqref{eq:df_def}, the QDF can be written as a single contour-ordered exponential
\be \Phi_{ \alpha \beta}(t) = \Braket{\tordc \left\{  e^{-i \intc  \mc{B}_{\alpha \beta}(\tau) \,d\tau} \right\}} ,
\label{eq:solq2}
\ee 
 where the contour integral is defined as 
 $ \intc  = \int_{0+i\eta}^{t+i\eta} - \int^{t-i\eta}_{0-i\eta}$.


\subsubsection{Cumulant expansion}

With Eqs. \eqref{eq:solq2} and \eqref{eq:solc},  the condition \eq{eq:cond} becomes  
\be\overline{e^{-i\int_0^t \Delta_{\alpha \beta}(s) \,ds }} = \Braket{\tordc e^{-i \intc \mc{B}_{\alpha\beta}(\tau) \,d\tau}} \label{eq:final}. 
\ee 
While formally exact, it is still nontrivial to directly infer from \eq{eq:final} whether it is possible to find random processes $ \{\eta_\alpha(t)\}$  that satisfy it. 
Further progress can be made 
 by performing a cumulant expansion for both sides of \eq{eq:final}, 
 
 \be
 \begin{split} 
 \ln \Phi_{\alpha \beta}(t) &\equiv K^\text{q}_{\alpha \beta}(t) = \sum_{n=1}^\infty \frac{(-i)^n}{n!}\kappa_{\alpha \beta}^{\text{q}, (n)}(t), \\ 
  \ln \Phi^\text{noise}_{\alpha \beta}(t) &\equiv K^\text{c}_{\alpha \beta}(t) = \sum_n \frac{(-i)^n}{n!}\kappa_{\alpha \beta}^{\text{c}, (n)}(t). 
\end{split} 
\ee 
The cumulant expansion is the Taylor expansion of the logarithm of the decoherence function with respect to the system-bath coupling strength. This can readily seen by parameterizing the system-bath interaction as $H_\text{SB} \rightarrow \lambda H_\text{SB}$. 

 For the classical and quantum decoherence functions to be equivalent irrespective of the system-bath interaction strength, the cumulants of $\Phi_{\alpha \beta}(t)$ and $\Phi^\text{noise}_{\alpha \beta}(t)$ need to match order by order. This condition is, in fact, stricter than \eq{eq:final}. 
For the NIDF, the cumulant expansion can be obtained through the following recursive formula \cite{Smith1995} 
\be
 \kappa^{\text{c}, (n)}_{\alpha \beta}  = \mu^{\text{c}, (n)}_{\alpha\beta} - \sum_{m=1}^{n-1} \binom{n-1}{m-1} \kappa^{\text{c}, (m)}_{\alpha \beta} \mu^{\text{c}, (n-m)}_{\alpha \beta},
\label{eq:recursive}
\ee
where 
\be \mu^{\text{c}, (n)}_{\alpha\beta} =  \idotsint_0^t \overline{\Delta_{\alpha \beta}(s_1) \cdots \Delta_{\alpha \beta}(s_n) }\, ds_1\cdots ds_n 
\label{eq:c_moments}
\ee 
are the moments of the stochastic variable $\Delta_{\alpha \beta}$ and $\binom{n}{m}$ denote the binomial coefficients. 
One of the advantages of recasting the quantum decoherence function into a single exponential  is that it becomes simpler to perform a cumulant expansion.  A straightforward extension of the cumulant expansion for time-ordered exponentials by Kubo \cite{Kubo1962} leads to  the conclusion that the quantum cumulants satisfy the same recursive formula \eq{eq:recursive}, that is, 
\be
\kappa^{\text{q}, (n)}_{\alpha \beta}  = \mu^{\text{q}, (n)}_{\alpha\beta} - \sum_{m=1}^{n-1} \binom{n-1}{m-1} \kappa^{\text{q}, (m)}_{\alpha \beta} \mu^{\text{q}, (n-m)}_{\alpha \beta},
\label{eq:q_recursive}
\ee
 with the generalized quantum moments of operator $\mc{B}_{\alpha \beta}$ defined as  
\be \mu_{\alpha \beta}^{\text{q}, (n)} = \idotsint_\mc{C} \Braket{\tordc \prod_{i=1}^n  \mc{B}_{\alpha \beta}(\tau_i)}  \,\prod_{i=1}^n d\tau_i. 
\label{eq:q_moments}
\ee

With the cumulant expansion for both sides of \eq{eq:final}, the problem of whether classical noise can mimic quantum pure-dephasing dynamics can now be mapped to the much more manageable task of whether one can find a classical noise having  correlation functions equivalent to the quantum time-correlation functions.

The first-order cumulant  of  the quantum and noise-induced decoherence function reads  
\be \kappa^\text{q,(1)} =  \intc d\tau \braket{\mc{B}_{\alpha \beta}(\tau)} = \int_0^t \braket{\tilde{B}_\alpha(s) - \tilde{B}_\beta(s)} \,ds  , 
\ee
\be  \kappa^\text{c,(1)} = \int_0^t  \overline{ \Delta_{\alpha \beta}(s)} \,ds = \int_0^t \overline{ \eta_{\alpha}(s)-\eta_{\beta}(s) } \,ds . \ee 
At a quantum level this cumulant is determined by the expectation value of the environment operators entering $H_\text{SB}$. At a noise level it is determined by the expectation value of the noise. Since 
the expectation value of the environment operator is merely a real number, 
it is always possible to find noise with its average  $\overline{\eta_\alpha(t)} = \braket{\tilde{B}_\alpha(t)}$ such that $\kappa^\text{q,(1)} = \kappa^\text{c,(1)}$.

   A more stringent requirement comes from the second cumulant. As it is always possible to redefine the system Hamiltonian to make the expectation value of the environment operator vanish, we assume that the first cumulant vanishes in the following. 
From Eqs. (\ref{eq:recursive} - \ref{eq:q_moments}), it is straightforward to obtain  the second cumulant for the QDF and NIDF
\be \kappa^{\text{c}, (2)}_{\alpha \beta} =  \iint_0^t dsds' \overline{\Delta_{\alpha \beta}(s)\Delta_{\alpha \beta}(s')},  
\label{eq:second_c}
 \ee 
and 
\be 
\begin{split} 
\kappa^{\text{q}, (2)}_{\alpha\beta}(t) =& \iint_\mc{C}d\tau d\tau' \Braket{\tordc \mc{B}_{\alpha \beta}(\tau) \mc{B}_{\alpha \beta}(\tau') } \\    
= &  2\int_0^t ds\int_0^s ds'  (D_{\alpha\alpha}(s,s') + D_{\beta \beta}(s',s)) \\ 
&-2\iint_0^t ds ds' D_{\beta \alpha}(s,s')
\end{split}
\label{eq:second_q}
\ee  
where $ D_{\alpha \beta}(s,s') = \Braket{ 	\tilde{B}_\alpha(s) \tilde{B}_\beta(s')}$ is the quantum  time-correlation function of the environment.  
Because  the classical noise is real, if the second cumulant for the QDF is complex, the classical noise cannot fully capture the effects of a quantum environment.  Thus, a necessary condition to mimic the quantum decoherence with classical noise is that  the cumulants are real.

Higher-order cumulants can be important for anharmonic and many-body environments.  Using \eq{eq:recursive}, it is now straightforward to obtain higher-order cumulants for QDF. For example, the  third cumulant is given by 
\be \kappa^{\text{q}, (3)}_{\alpha \beta} = \iiint_\mc{C} \braket{\tordc \mc{B}_{\alpha \beta}(\tau_1) \mc{B}_{\alpha \beta}(\tau_2)\mc{B}_{\alpha \beta}(\tau_3)} \,d\tau_1d\tau_2 d\tau_3. 
\ee 
If the higher-order quantum cumulants make significant contributions to decoherence, it requires the classical noise to have the corresponding higher-order correlations. This implies that, for such environments,  the commonly used Gaussian noise model can be inadequate \cite{Yang2017, Kubo1963}.   We expect that such environments can arise in electronic decoherence in molecules where the environment are molecular vibrations which can be far from harmonic, and also in central spin model  where the environment consists of interacting spins.

Surprisingly, the cumulants, often considered as a convenient computational tool, carry direct physical meaning. To see this, we take a time-derivative of \eq{eq:solq} and use the definition of the cumulants to obtain  
\be \frac{d}{dt} {\tilde{\rho}}^\text{S}_{\alpha\beta}(t) = \dot{K}^\text{q}_{\alpha\beta}(t) \tilde{\rho}^\text{S}_{\alpha\beta}(t)
\label{eq:122}
\ee 
Equation \eqref{eq:122} is the equation of motion for the coherences in the interaction picture. Clearly, the time-derivative of the cumulants are the generators of decoherence and each cumulant corresponds to  a particular order on the system-bath interaction. Explicitly, expressing the coherence in the polar form ${\tilde{\rho}}^\text{S}_{\alpha\beta}(t) = A_{\alpha \beta}(t)e^{i\phi_{\alpha \beta}(t)}$, it follows from \eq{eq:122} that 
\be \dot{A}_{\alpha\beta}(t) = \Re \dot{K}^\text{q}_{\alpha\beta}(t) {A_{\alpha\beta}(t)}, ~~~ \dot{\phi}_{\alpha \beta}(t) = \Im \dot{K}^\text{q}_{\alpha\beta}(t). 
\label{eq:124}
\ee 
 Equation \eqref{eq:124} indicates  that the real parts of the time-derivative of cumulants is responsible for decoherence, and the imaginary parts account for the  environment-induced energy shifts. 

\subsection{Spin-boson model}
We now illustrate how the above criteria can be applied using a concrete example: the quintessential spin-boson problem. 
 The Hamiltonian for the pure-dephasing spin-boson model is 
\be H = - \frac{\omega_0}{2}\sigma_z + \sigma_z \sum_k g_k (a_k^\dag + a_k) + H_\text{B}. 
\label{eq:h_sb}
\ee
where $\sigma_z$ is the Pauli $z$ matrix and $\omega_0$ the transition frequency for the two-level system. Here  $H_\text{B} = \sum_k \omega_k a_k^\dag a_k$ describes a bosonic environment consisting of  a distribution of harmonic oscillators of frequency $\omega_k$  with $a_k, a^\dag_k$ being the creation and annihilation operators for the $k$-th mode, respectively.  The coupling of the system with the environment leads to shifts in the system's energy levels, where $g_k$ is the coupling constant to the $k$-th harmonic mode.

The  environment is assumed to be initially in  thermal equilibrium at inverse temperature $\beta = 1/(k_\text{B}T)$ with density matrix $\rho_\text{B} = e^{-\beta H_\text{B}}/Z $ where $Z = \tr_\text{B}[e^{-\beta H_\text{B}}]$ is the partition function. For time-independent Hamiltonian, this leads to  time-translational invariant time-correlation function 
\be D_{\alpha\beta}(t,t') = D_{\alpha\beta}(t-t') .\ee
For a two-level system, only one  decoherence function has to be considered corresponding to $\alpha = 0, \beta = 1$. Since $\sigma_z = \ket{0}\bra{0} -\ket{1}\bra{1}$, one  can identify $B_0 = - B_1 = \sum_k g_k (a_k + a^\dag_k) \equiv B$ and $D_{00} = D_{11} = -D_{01} \equiv D $.

Using $\tilde{a}_k(t) = e^{-i\omega_k t}a_k$ and $\tilde{a}^\dag_k(t) = e^{i\omega_k t}a_k^\dag$,
the time-correlation function $D(t)$ can be calculated as 
\be 
\begin{split} 
D(t) &= \sum_k |g_k|^2 \left( \Braket{\tilde{a}_k(t) a_k^\dag} + \Braket{\tilde{a}_k^\dag(t)a_k} \right) \\ 
&=  \sum_k |g_k|^2 [(1 - \bar{n}_k) e^{-i\omega_k t} + \bar{n}_k e^{i\omega_k t}] 
\end{split}
\label{eq:tmp111} 
\ee 
where $\bar{n}_k = \braket{a_k^\dag a_k}$ is the distribution function. 
At thermal equilibrium, $\bar{n}_k = 1/(e^{\beta \omega_k} -1)$ corresponding to the Bose-Einstein distribution and \eq{eq:tmp111} yields 
\be 
\begin{split} 
D(t) &=  \int_0^\infty  \frac{d\omega}{\pi} J(\omega) [\coth(\beta \omega/2 ) \cos(\omega t) - i\sin(\omega t) ] \\ 
&= \intf  \frac{d\omega}{2\pi} J(\omega) [\coth(\beta \omega/2 ) \cos(\omega t) - i\sin(\omega t) ]
\end{split} 
\label{eq:tcf}
\ee 
where the spectral density is  defined as $ J(\omega) \equiv  \pi \sum_k |g_k|^2 \delta(\omega - \omega_k)$ for $\omega > 0$  and extended to negative frequencies by   $J(-\omega) = -J(\omega)$. This extension makes the integrand in \eq{eq:tcf} symmetric under $\omega \rightarrow -\omega$, hence the second equality. 
Since the environment is Gaussian, the QDF is determined by the first two cumulants \cite{Kubo1962, Wick1950}. The first cumulant vanishes, and   
the second cumulant can be calculated by inserting \eq{eq:tcf} into \eq{eq:second_q}
\be \kappa_{\alpha \beta}^{\text{q}, (2)}(t) = 8 \intf \frac{d\omega}{2\pi} J(\omega) \coth(\beta \omega/2) \frac{1-\cos(\omega t)}{\omega^2} 
\label{eq:sb_DF}
\ee 
Interestingly, the cumulant is real even though the time-correlation function is complex. This is due to the property of the quantum time-correlation function 
\be D(-\tau) = D^*(\tau). 
\label{eq:sym}
\ee 
Because the cumulant is real, as described below, its effects on the dynamics can be mimicked by classical noise.

Consider now the noise model intended to mimic the above decoherence dynamics  with Hamiltonian 
\be 
H(t) = -\frac{\omega_0}{2}\sigma_z + \eta(t) \sigma_z.
\ee 
where the stochastic process $\eta(t)$ replaces the system-bath interaction in \eq{eq:h_sb}. 
Denoting the noise correlation function as $C(s,s') = \overline{\eta(s)\eta(s')}$, we show that if the noise satisfy the following three conditions: (i) $C(s,s') = C(s-s')$, (ii) $\overline{\eta(t)} = 0$, and (iii) $C(t) = S(t)$ where $S(t) \equiv \half \braket{\{B(t), B\}} = \half \braket{B(t)B + BB(t)}$, then the NIDF coincides with the QDF. The first condition implies that the noise is stationary corresponding to the equilibrium state of the environment. The second condition reflects the vanishing of the first cumulant of the QDF. The third one is required to make the second cumulants for QDF and NIDF equal. To see this, realizing that $\Delta_{01}(t) = 2\eta(t)$ and inserting the Fourier transform of the noise correlation function 
\be C(t) = \intf \frac{d\omega}{2\pi} C(\omega) e^{- i\omega t} , \ee
 into \eq{eq:second_c} yields   
\be \kappa^{\text{c}, (2)}(t) = 8\intf \frac{d\omega}{2\pi}  \frac{1-\cos(\omega t)}{\omega^2} C(\omega)
\label{eq:117}
\ee 
Comparing  \eq{eq:sb_DF} and \eq{eq:117}, it is clear that the condition  $\kappa^{\text{q}, (2)}(t) = \kappa^{\text{c}, (2)}(t)$ is equivalent to 
\be  {C}(\omega) = J(\omega) \coth(\beta\omega/2). \label{eq:cond3} 
\ee
According to  \eq{eq:tcf},  the right-hand side of \eq{eq:cond3} is the Fourier transform of  the real part of the quantum time-correlation function [\eq{eq:tcf}]. 
Using \eq{eq:sym}, it follows  that
$ S(t) = \Re D(t) $
and thus to the third condition $S(t) = (1/2)(D(t) + D(-t)) = (1/2) \langle \{B(t), B\}
\rangle$.  

Equation \eqref{eq:cond3} suggests that for each spectral density there is a corresponding classical noise leading to the same pure-dephasing dynamics provided that an adequate algorithm to
generate the  stochastic process is identified.  
Here we exemplify the analysis with the widely used 
 Ohmic environments with a Lorentz-Drude cutoff. The spectral density for such environments is   
\be J(\omega) = 2\lambda  \frac{\omega_c \omega}{\omega^2 + \omega_c^2}, \ee  
where $\omega_c$ is the cutoff frequency of the environment and $\lambda$ characterizes the system-bath interaction strength.  
  In the high-temperature limit  $\beta \omega_c \ll 1$, $\coth(\beta \omega/2) \approx 2(\beta \omega)^{-1}$ and
\be J(\omega) \coth(\beta \omega/2) \approx  4\lambda k_\text{B} T \frac{\omega_c}{\omega^2 + \omega_c^2}
\label{eq:sd}
\ee 
Now  let $\eta(t)$ be a colored Gaussian noise with correlation function $C(\tau) = 2 \lambda k_\text{B} T e^{-\omega_c \tau}$. This choice ensures that \eq{eq:cond3} is satisfied in the high temperature limit which can be seen by taking the Fourier transform of the noise correlation function and comparing with \eq{eq:sd}. Therefore, the quantum pure-dephasing effects of a high-temperature Ohmic bath can be fully captured by colored exponentially correlated Gaussian noise. 

 \begin{figure}[htbp]
	\includegraphics[width=0.5\textwidth]{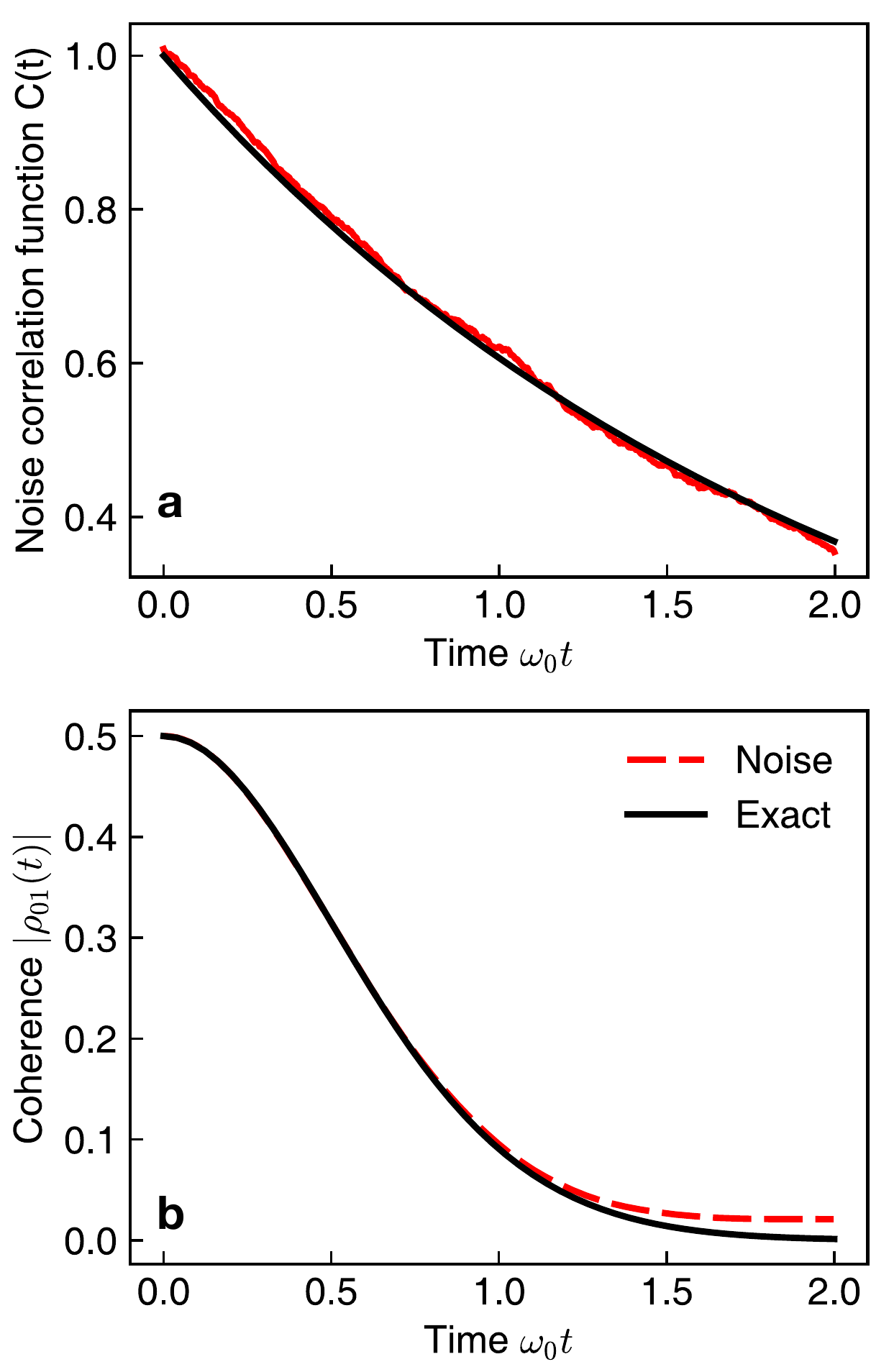}
	\caption{(a) Correlation function of the generated noise (red) in comparison to the target (black). (b) Quantum and noise-induced decoherence dynamics in a spin-boson model starting from a superposition with equal coefficients of ground and excited state.    Model parameters are: $\lambda/\omega_0 = 0.5, \beta \omega_0 = 1, \omega_c/\omega_0 = 1$. The exact results is obtained through $\Phi_{01}(t) = e^{-\half \kappa^{\text{q}, (2)}(t)}$ with the second cumulant computed using \eq{eq:sb_DF}. The stochastic simulations are obtained with 2000 realizations of the colored noise and with a time step $\omega_0 dt = 0.002$. No revivals of the coherence are observed in this model. 
} 
	\label{fig:spin_boson} 
\end{figure}

 This conclusion is demonstrated in \fig{fig:spin_boson}, which contrasts the exact quantum results with stochastic simulations.  The exact results are obtained by first inserting  \eq{eq:sd} into \eq{eq:sb_DF} to obtain the second-order cumulant and thus the decoherence function $\Phi_{01}(t) = e^{-\half \kappa^{\text{q}, (2)}(t)}$.   This decoherence function is exact (compare with Ref.~\onlinecite{Breuer2002b}) as the contributions of higher order cumulants vanish in this case. 
 The stochastic simulation is averaged over 2000 realizations of the  exponentially correlated colored Gaussian noise generated using the algorithm in \cite{Fox1988}. The  correlation function of generated noise is shown in Fig. \ref{fig:spin_boson}a. For each realization, the stochastic \tdse\ $i \frac{d}{dt} \ket{\psi(t)} = H(t)\ket{\psi(t)}$ with the initial condition $\ket{\psi(0)} = \frac{1}{\sqrt{2}} (\ket{0} + \ket{1})$ is integrated. As shown, the  decoherence dynamics obtained with stochastic noise is in quantitative agreement with the exact quantum decoherence dynamics, consistent with our conclusion above. 

For low-temperature regime and other types of spectral densities, if $S(t)$ can be well-described by a set of exponential functions,
\be S(t) =  \sum_n |c_n|^2 e^{- |t-t'|/\tau_n}, \ee
one can choose a sum of exponentially-colored Gaussian noises
\be \eta(t) =  \sum_n c_n \eta_n(t) \ee 
 where $\{\eta_n(t)\}$ are  Gaussian stochastic processes with statistical properties  
 \be \overline{\eta_n(t)\eta_m^*(t')} = \delta_{nm} e^{- |t-t'|/\tau_n}. \ee 
In this case, the noise correlation function 
\be C(t) = \sum_{n, m} c_n c_m^* \overline{\eta_n(t)\eta_m^*(t')} = \sum_n |c_n|^2 e^{- |t-t'|/\tau_n} = S(t). 
\ee
 Thus, the quantum decoherence dynamics can still be captured by classical noise.

 \section{Quantum dissipation} \label{sec:diss}
  Another major source of decoherence is quantum dissipation due to { transitions} between system eigenstates induced by the environment.  The role of the dissipative environment is to  drive the system from  an initially out-of-equilibrium state to 
 thermal equilibrium.  
 
 The question we seek to address here is when can we understand quantum decoherence induced by dissipation in terms of classical noise.  This problem has been studied previously by Tanimura and Kubo \cite{Tanimura1989} with the hierarchical equation of motion. The conclusion of such a formal study is that the classical noise can only be made to be equivalent to a full quantum treatment at infinite temperature, i.e., as $\beta \rightarrow 0$ . Below we provide a simpler analysis of this problem for Markovian environments and show that the physical reason behind this conclusion is that the classical noise cannot describe the decoherence effects due to spontaneous emission induced by a dissipative environment. Here spontaneous emission is not restricted to electromagnetic environments but refers to a damping effect induced by the spontaneous fluctuations of any dissipative environment. 
 
 The simplest model that allows  isolating this basic physics is a two-level system $\ket{g}, \ket{e}$ interacting with a thermal  environment. A standard full quantum treatment of this model within the dipole approximation  
 leads to the equation of motion for the reduced density matrix \cite{Breuer2002a} 
 \be 
 \begin{split}
 	\dt {\rho}_{\text{S}}(t) =& -i [H_\text{S}, \rho_\text{S}] +  \Gamma_e \left( \sigma_- \rho_\text{S}(t)\sigma_+ - \half \{ \sigma_+\sigma_- , \rho_\text{S}(t)\}\right)\\
 	& + \Gamma_a \left( \sigma_+ \rho_\text{S}(t)\sigma_- - \half \{ \sigma_-\sigma_+ , \rho_\text{S}(t)\}\right)
 \end{split}
 \label{eq:bloch}
 \ee 
 where $H_\text{S}= -\omega_0 \sigma_z/2$ is the system Hamiltonian and $\sigma_\pm$ is the raising/lowering operator, $[A,B] = AB - BA$ and $\{A, B\} = AB + BA$ denote the commutator and anticommutator, respectively.  The first term in the right-hand side of \eq{eq:bloch} accounts for the unitary dynamics of $H_\text{S}$, which does not contribute to decoherence.  
 The meaning of the remaining dissipative terms is best revealed by decomposing \eq{eq:bloch} in terms of the matrix elements 
 \be \dt \rho_{gg}^\text{S}(t) =  \Gamma_e \rho_{ee}^\text{S}(t) - \Gamma_a \rho_{gg}^\text{S}(t)
 , \ee
 \be 
 \dt \rho_{eg}^\text{S}(t) = -i \omega_0 \rho_{eg}^\text{S}(t) - \Gamma_\text{d} \rho_{eg}^\text{S}(t). 
 \ee  
 where $\Gamma_\text{d}  = (\Gamma_e + \Gamma_a)/2$. 
 Clearly, the second term in \eq{eq:bloch} accounts for the emission of energy
 to the environment, and the third one to absorption. 
Here the emission rate $\Gamma_{e}$  is a sum of the stimulated emission rate (which is equivalent
 to the absorption rate  $\Gamma_{a}$) and spontaneous emission rate $\Gamma_0$, i.e.,  $\Gamma_e = \Gamma_a + \Gamma_0$. The off-diagonal matrix elements (or coherence) represented in the eigenstates of $H_\text{S}$ admits an exponential decay with the decoherence rate $\Gamma_\text{d}$.

 Note that as a consequence of the Markovian approximation involved in the derivation of \eq{eq:bloch}, the model does not capture  the universal
 initial Gaussian regime for uncorrelated initial states which gives rise to quantum Zeno effects \cite{Fischer2001, Facchi2004, Gu2018b}.

 Now consider the classical noise picture where the system is subject to a random term that induces  transitions between system eigenstates, i.e., 
 \be H = H_\text{S} + \eta(t)\sigma_- + \eta^*(t) \sigma_+ 
 .\ee 
   Here the stochastic variable $\eta$ are allowed to be complex but still keeping the dynamics for each noise realization unitary. 
 For Markovian environments without memory effects, it is appropriate to choose $\braket{\eta(t)\eta^*(t')} = \gamma  \delta(t-t')$.  In the interaction picture of $ H_\text{S}$, the Liouville von-Neumann equation reads 
 \be i \dt \tilde{\rho}_\text{S}(t) = [\eta(t)\tilde{\sigma}_-(t) + \eta^*(t) \tilde{\sigma}_+(t) , \tilde{\rho}_\text{S}(t)] 
 \label{eq:lvn}
 \ee 
 
 A quantum master equation can be obtained as follows. Integrating \eq{eq:lvn} yields 
 \be \tilde{\rho}_\text{S}(t) = \rho_{\text{S}}(0) - i \int_0^t dt' [\eta(t')\tilde{\sigma}_-(t') + \eta^*(t') \tilde{\sigma}_+(t') , \tilde{\rho}_\text{S}(t')].
 \label{eq:120}
 \ee  
 Inserting \eq{eq:120} back into the right-hand side of \eq{eq:lvn} and taking statistical average of the stochastic processes yields 
 \be 
 \begin{split} 
 	\frac{d}{dt} \tilde{\rho}_\text{S}(t) =& \gamma [ \tilde{\sigma}_-(t), [\tilde{\sigma}_+(t), \tilde{\rho}_\text{S}(t)]] + \gamma [ \tilde{\sigma}_+(t) , [\tilde{\sigma}_-(t) , \tilde{\rho}_\text{S}(t)]]. \\ 
 \end{split}
 \label{eq:123}
 \ee 
 Transforming into the Schr\"{o}dinger picture gives the quantum master equation 
 \be 
 \begin{split} 
 	\dot{\rho}_\text{S}(t) =&  -i [H_\text{S}, \rho_\text{S}(t)] +  \gamma \left( \sigma_- \rho_\text{S}(t)\sigma_+ - \half \{ \sigma_+\sigma_- , \rho_\text{S}(t)\}\right) \\
 	&+ \gamma \left( \sigma_+ \rho_\text{S}(t)\sigma_- - \half \{ \sigma_-\sigma_+ , \rho_\text{S}(t)  \} \right). 
 \end{split}
 \label{eq:111}
 \ee 
 Comparing \eq{eq:111} and \eq{eq:bloch}, it becomes clear that the noise can mimic many of the effects of the quantum relaxation provided that one identifies $\gamma$ with $\Gamma_a$.  What becomes missing in this picture are the contributions due to spontaneous emission. In this case, one obtains a decoherence rate $\gamma_\text{d} = \Gamma_a$ from \eq{eq:111}. Thus, the decoherence rate in the classical noise picture does not contain the contribution from spontaneous emission. 

The missing of spontaneous emission has a direct consequence in relaxation. Since the absorption and emission  rates are equal, the stationary state at long times is the non-physical infinite-temperature state. This problem can be fixed by going beyond the classical noise picture.  For example, by promoting the classical noise to quantum noise \cite{Gardiner2004} or by relaxing the constraint of unitary dynamics for each noise realization as in the stochastic Liouville equation \cite{Hsieh2018}.

\section{Conclusions} \label{sec:conc} 
To summarize, we have contrasted quantum decoherence that arises as a single quantum system becomes entangled with environmental degrees of freedom with the apparent decoherence that results by averaging over an ensemble of unitary evolutions generated by a Hamiltonian subject to classical noise. For dissipative environments, we showed that the classical noise cannot describe the decoherence induced by spontaneous emission and, thus, that the classical noise picture can only become quantitative in the infinite temperature limit. For pure-dephasing dynamics, we identified general conditions that determine whether the decoherence dynamics due to a quantum environment can be quantitatively mimicked through classical noise. Specifically, we showed that for the two dynamics to agree the cumulants of the quantum and noise-induced decoherence functions must coincide. These requirements impose restrictions on the statistical properties of the noise that are determined by the quantum  many-point time correlation function of the environmental operators that enter into the system-bath interaction.
 These conditions are valid for {any} pure dephasing problem including anharmonic environments and nonlinear system-bath couplings. 
 
 In particular, through the spin-boson model, we demonstrated numerically and analytically  that  the decoherence effects due to a harmonic Ohmic environment (in the  high-temperature pure-dephasing  limit) can be mimicked by exponentially correlated colored Gaussian noise. This observation is consistent with a recent study~\cite{Rahman2019} of the quantum transport properties of a molecular junction subject to vibrational dephasing that finds agreement between a fully quantum model (harmonic, Ohmic, pure-dephasing environment in the  high temperature limit) and a model in which the thermal environment manifests itself in (exponentially correlated Gaussian) fluctuating  site energies. A challenge in employing classical noise models for environments with more complicated spectral densities is to generate noise with the correct statistical properties. 
 
The classical noise model has also been  useful in quantum information processing \cite{Rossi2014}, particularly for the design of dynamic decoupling schemes to preserve coherence \cite{Yang2017}. In particular, in the context of optimal control computations an effective stochastic model that captures the effects of a quantum environment is highly desirable \cite{Witzel2014} as these computations are challenging for a full quantum model. Our results offers well-defined criteria to develop  and to understand the limitations of such models.  

\begin{acknowledgments}
	This material is based upon work supported by the National
	Science Foundation under CHE-1553939.	
\end{acknowledgments}

\bibliography{coherence}

\begin{thebibliography}{47}
\bibitem [{\citenamefont {Gu}\ and\ \citenamefont {Franco}()}]{Gu2018b}%
  \BibitemOpen
  \bibfield  {author} {\bibinfo {author} {\bibfnamefont {Bing}\ \bibnamefont
  {Gu}}\ and\ \bibinfo {author} {\bibfnamefont {Ignacio}\ \bibnamefont
  {Franco}},\ }\bibfield  {title} {\enquote {\bibinfo {title} {Consequences of
  microscopic reversibility and early-time expansions on quantum decoherence
  dynamics},}\ }\href@noop {} {\bibinfo  {journal} {To be Summitted}\
  }\BibitemShut {NoStop}%
\end{thebibliography}
\end{document}